\documentclass{article}

\usepackage{arxiv}

\usepackage[utf8]{inputenc} 
\usepackage[T1]{fontenc}    
\usepackage{url}            
\usepackage{booktabs}       
\usepackage{amsfonts}       
\usepackage{nicefrac}       
\usepackage{microtype}      
\usepackage{lipsum}		
\usepackage{graphicx}
\usepackage{natbib}
\usepackage{doi}

\usepackage{graphicx}
\usepackage{multicol,multirow}
\usepackage{amsmath,amssymb,amsfonts}
\usepackage{mathrsfs}
\usepackage{mathalpha}
\usepackage{amsthm}
\usepackage{rotating}
\usepackage{appendix}
\usepackage{ifpdf}
\usepackage[T1]{fontenc}
\usepackage{newtxtext}
\usepackage{newtxmath}
\usepackage{textcomp}
\usepackage{xcolor}
\usepackage{lipsum}

\title{Meta-Analysis with JASP, Part I: Classical Approaches}

\author{ \href{https://orcid.org/0000-0002-0018-5573}{\includegraphics[scale=0.06]{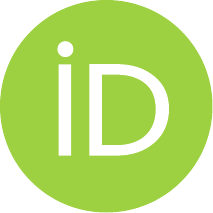}\hspace{1mm}František Bartoš} \\
	Department of Psychological Methods\\
	University of Amsterdam\\
	\texttt{f.bartos96@gmail.com} \\
	\And 
        \href{https://orcid.org/0000-0003-1596-1034}{\includegraphics[scale=0.06]{orcid.pdf}\hspace{1mm}Eric-Jan Wagenmakers} \\
	Department of Psychological Methods\\
	University of Amsterdam\\
	\And
	\href{https://orcid.org/0000-0003-3463-4063}{\includegraphics[scale=0.06]{orcid.pdf}\hspace{1mm}Wolfgang Viechtbauer} \\
	Department of Psychiatry and Neuropsychology\\
	Maastricht University\\ \\
}

\hypersetup{
 pdftitle    = {Meta-Analysis with JASP, Part I: Classical Approaches},
 pdfsubject  = {stat.ME},
 pdfauthor   = {František Bartoš, Eric-Jan Wagenmakers, Wolfgang Viechtbauer},
 pdfkeywords = {Open-Source Software, JASP, Meta-Analysis},
}

\begin{document}
\maketitle

\begin{abstract}
Meta-analyses play a crucial part in empirical science, enabling researchers to synthesize evidence across studies and draw more precise and generalizable conclusions. Despite their importance, access to advanced meta-analytic methodology is often limited to scientists and students with considerable expertise in computer programming. To lower the barrier for adoption, we have developed the Meta-Analysis module in JASP (\url{https://jasp-stats.org/}), a free and open-source software for statistical analyses. The module offers standard and advanced meta-analytic techniques through an easy-to-use graphical user interface (GUI), allowing researchers with diverse technical backgrounds to conduct state-of-the-art analyses. This manuscript presents an overview of the meta-analytic tools implemented in the module and showcases how JASP supports a meta-analytic practice that is rigorous, relevant, and reproducible. Tutorial videos accompany the examples presented in this manuscript.
\end{abstract}


\section{Introduction}

Meta-analyses are central to modern empirical science. In parallel, meta-analytic methodology has advanced rapidly, with many methods disseminated as packages in the \texttt{R} programming language (\cite{R}; see \url{https://cran.r-project.org/view=MetaAnalysis}). While this ecosystem is convenient for programming-savvy methodologists, many applied scientists (and their students) prefer not to analyze data using a programming language.

As an alternative, applied scientists often turn to `point-and-click' statistical software with a graphical user interface. For meta-analysis, prominent options include both proprietary (e.g., CMA \citep{CMA4}, RevMan \citep{RevMan}) and open-source (e.g., MetaWin \citep{metawin3}, MetaEssentials \citep{MetaEssentials}, OpenMEE \citep{OpenMEE}) software. However, these packages rarely provide advanced capabilities such as multilevel and multivariate models with complex dependency structures \citep{olkin2009stochastically, konstantopoulos2011fixed} or location--scale meta-regression \citep{viechtbauer2022location}. Many of these packages also lack comprehensive diagnostics and integration with interpretation tools such as estimated marginal means, contrasts, and publication-ready visualizations.

In order to overcome these limitations, we have developed the JASP Meta-Analysis module; this module provides a graphical user interface for the \texttt{metafor} \citep{metafor} \texttt{R} package enhanced with additional post-processing, integrated with the free and open-source statistical program JASP \citep{JASP95}. As such, the module provides a user-friendly graphical user interface to state-of-the-art meta-analytic tools. This manuscript focuses on the implemented statistical procedures rather than on detailed guidance about when to use them; for that, we refer readers to the cited literature and meta-analytic textbooks (see e.g., \cite{borenstein2009introduction, schmid2020handbook, borenstein2022systematic, harrer2021doing}).

In the following sections, we first briefly introduce JASP. Second, we demonstrate the meta-analytic methodology implemented in the Meta-Analysis module through three examples: Example~1 features the Bacillus Calmette-Guerin (BCG) vaccine and covers effect size computation, funnel plots, a random-effects meta-analysis, forest plots, and subgroup analysis; Example~2 features writing-to-learn interventions and illustrates meta-regression, bubble plots, estimated marginal means and contrasts, and location--scale models; finally, Example~3 features recidivism and mental health and demonstrates multilevel models, variance--covariance construction for multivariate models, and cluster-robust standard errors. Along the way, we highlight additional features and options available in the software.

The Meta-Analysis module also includes other functionality not treated here (e.g., meta-analyses of prediction model performance, publication bias adjustment, and meta-analytic SEM). A tutorial-style introduction of the Bayesian functionality is presented in Part II \citep{bartos2025bayesian}. JASP is available at \url{https://jasp-stats.org/download/}; source code can be found at \url{https://github.com/jasp-stats/}. Annotated JASP files are available on the OSF project page at \url{https://osf.io/mb56t/}. Tutorial videos are available on the JASP YouTube channel at \url{https://www.youtube.com/playlist?list=PLEIybnA0SEf0}.

\section{Introduction to JASP}

JASP is a free and open-source statistical software package whose development started in 2013 \citep{love2019JASP}. With over 100{,}000 downloads per month, JASP is used in the curricula of 389 universities from 77 different countries \citep[see also][]{field2025discovering}. JASP is primarily developed at the University of Amsterdam, with growing contributions from other institutions across the world (\url{https://jasp-stats.org/community-institutional-members/}).

Key features of JASP include a user-friendly point-and-click interface, dynamically updated result, progressive disclosure, translations into many languages, Open Science Framework  (OSF) integration, cross-platform support (Windows, macOS, Linux), an annotated data library, fully integrated AI support, and APA-style output with figures and tables exportable to Word or \LaTeX. The analysis output can be annotated and exported as an .html or .pdf file. The analysis itself can be saved as a .JASP file which allows users to share the analyses. The analysis files are reproducible in the same version of JASP (in case of analyses with sampling/bootstrap a seed needs to be set) and when uploaded to OSF the results can be directly previewed online. JASP runs locally, does not collect or send user information, and uses a sandboxed \texttt{R} environment, which guarantees that data always stay on one's computer (in contrast to online-hosted applications). 

Many of the analysis methods in JASP are accompanied by annotated example datasets and extensive documentation, accessible via on-hover tooltips or the help files linked through the `(i)' button in each analysis. Feature requests, bug reports, and questions are handled on the active GitHub issue tracker (\url{https://github.com/jasp-stats/jasp-issues}).

To replicate the examples that follow, you may download the current version of JASP from \url{https://jasp-stats.org/download/} and enable the Meta-Analysis module from the module panel using the `+' button in the top right corner of the application. To use the most up-to-date version of the Meta-Analysis module, you might need to update the module via the Module Store that opens under the same menu. Raw datasets and annotated JASP files for the examples below are included in the JASP library and at \url{https://osf.io/mb56t/}. Datasets can be loaded into JASP via the main menu in the top left corner of the application (`$\equiv$' button) and the `Open' option. JASP can import many different data formats, ranging from common spreadsheet formats like .csv, .xlsx to SPSS, Stata, SAS, Minitab, and R specific data formats. For a general overview of JASP, see \citep{wagenmakers2023accessible} and \citep{ly2021bayesian}.

\section{Example 1: Bacillus Calmette-Guerin (BCG) Vaccine}

The first example features the famous Bacillus Calmette-Guerin (BCG) vaccine dataset distributed with the \texttt{metadat} \texttt{R} package \citep{metadat}. Based on \citep{colditz1994efficacy}, the dataset features the counts of positive and negative tuberculosis cases for control (non-vaccinated) and experimental (vaccinated) groups across thirteen studies, along with absolute latitude and the treatment allocation procedure.

\subsection{Effect Size Computation}

\begin{figure}
    \centering
    \includegraphics[width=1\linewidth]{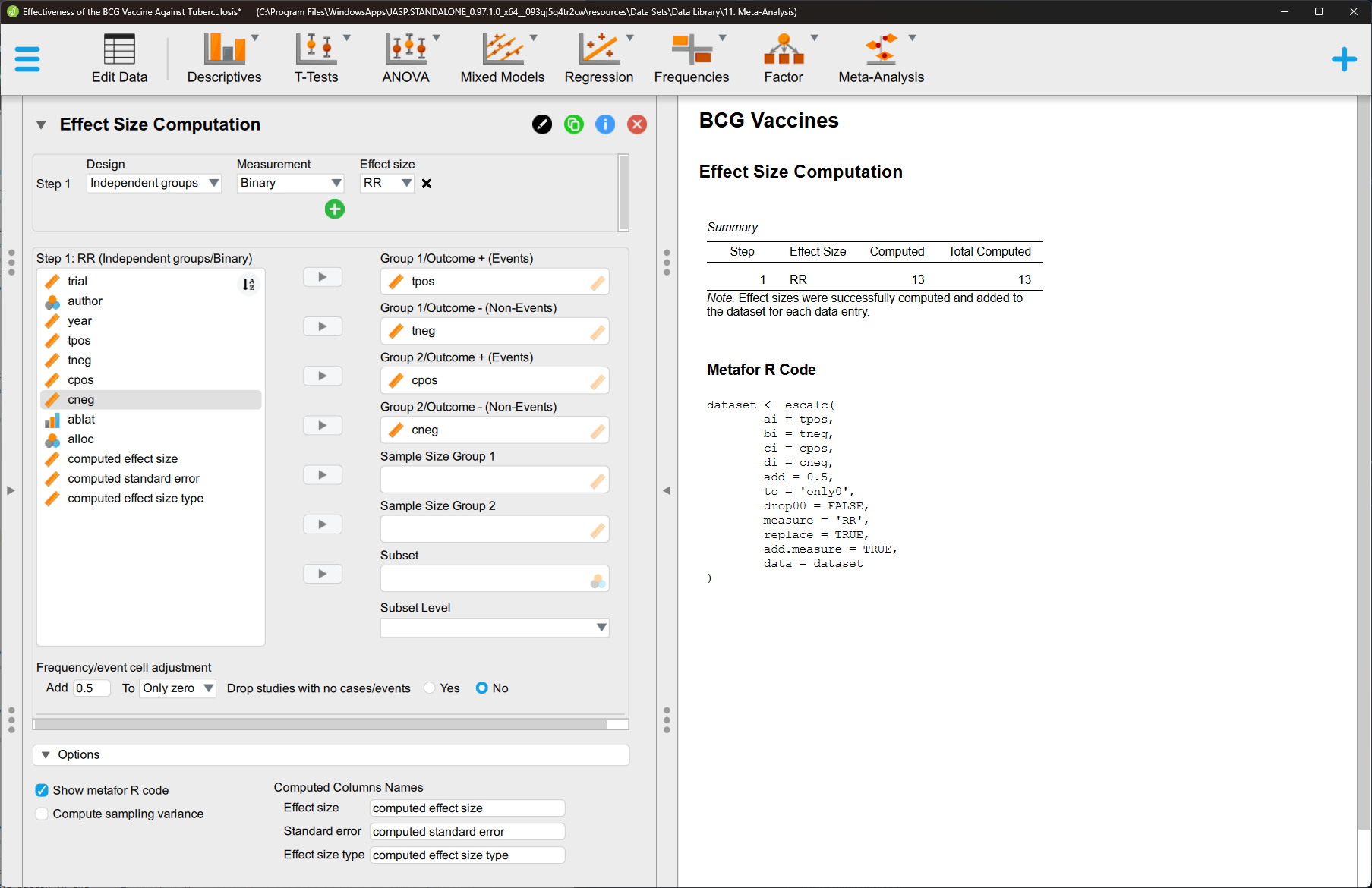}
    \caption{Effect Size Computation Menu in JASP}
    \label{fig:effect-size}
\end{figure}

We begin by loading the csv dataset (available from \url{https://osf.io/mb56t/}) into JASP via the $\equiv$' button in top left corner of the application and the `Open' option. Once the dataset is loaded into JASP, we proceed by computing effect size estimates and their standard errors from the case counts. This is done in the `Effect Size Computation' menu underneath the `Meta-Analysis' module heading (Figure~\ref{fig:effect-size}). In the left input panel, the `Design', `Measurement', and `Effect size' dropdowns determine which effect size is computed, and the variable inputs (e.g., `Group 1/Outcome +', `Group 1/Outcome -') map dataset columns to the calculations. The right output panel then shows how many effect sizes were computed successfully. Notice that the output panel dynamically updates as the options are selected; `run' or `execute' commands are usually not needed in JASP.

Behind the scenes, `Effect Size Computation' calls the \texttt{metafor::escalc} function. As such, the analysis supports a wide range of effect sizes suitable for many different settings. The corresponding \texttt{R} code can be generated using the `Show metafor R code' option in the `Options' section. Furthermore, the analysis allows for the chaining of multiple calculations with the green `+' button, and can restrict computations to subsets of rows via the corresponding `Subset' variable input. Details about the available effect sizes and the required variable input are available through the `(i)' help button in the top-right of the input section.

\subsection{Funnel Plot}

Next, the data are visualized with a funnel plot. Funnel plots are a useful tool for visual inspection of the data before proceeding with the analyses; for instance, funnel plots can help identify odd patterns, incorrectly extracted information (e.g., standard errors mistaken for standard deviations), and are used to examine the data for `small-study effects'.

\begin{figure}
    \centering
    \includegraphics[width=0.75\linewidth]{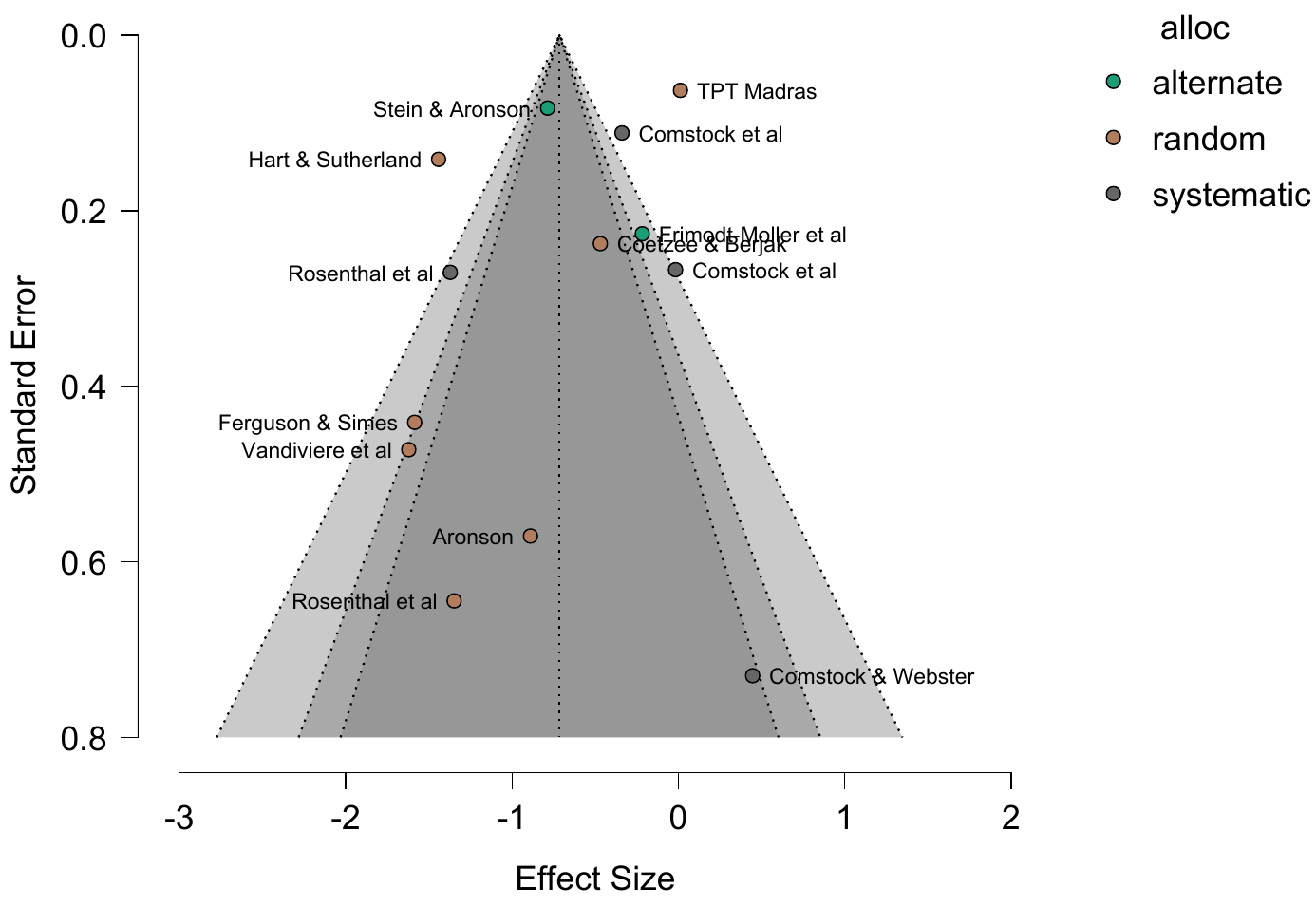}
    \caption{Funnel Plot Created in JASP}
    \label{fig:funnel-plot}
\end{figure}

The funnel plot is created using the `Funnel Plot' menu. The analysis requires specification of the computed `Effect Size' and `Effect Size Standard Error' variable inputs. By default, JASP shows a scatterplot of the effect sizes against their standard errors with a funnel that reflects the sampling distribution centered at the random-effects estimate (Figure~\ref{fig:funnel-plot}; a second funnel under the null hypothesis can be created via the `$\mathcal{H}_0$' checkbox). In the `Estimates Mapping' subsection, the `Study Label' option is set to the `author' variable and the `Color' option is set to the `allocation' variable, which extends the funnel plot with the author labels and allocation color coding. The figure can be exported in various file formats via the small triangle just above the plot.

Additional options include incorporating heterogeneity into the sampling distribution, displaying power enhancement (``power bands'') \citep{kossmeier2020power}, and creating separate funnels for subsets of data. Related statistical tests are in the `Publication Bias / Sensitivity Analyses' section, including funnel plot asymmetry tests \citep{egger1997bias, begg1994operating}, trim-and-fill \citep{duval2000trim}, and fail-safe N variants \citep{rosenthal1979file, orwin1983failsafe, rosenberg2005filedrawer}.

\subsection{Meta-Analysis}

We proceed by estimating a random-effects model via the `Meta-Analysis' menu in the `Classical' part of the `Meta-Analysis'. The user interface contains several sections separating it into easy-to-navigate option groups. In the pre-opened data input section, we specify the required `Effect Size' and `Effect Size Standard Error' variable inputs. Default settings follow \texttt{metafor} recommendations, including the `REML' random-effects estimator and the `Knapp and Hartung' adjustment for the fixed-effects test \citep{knapp2003improved}.

\begin{figure}
    \centering
    \includegraphics[width=1\linewidth]{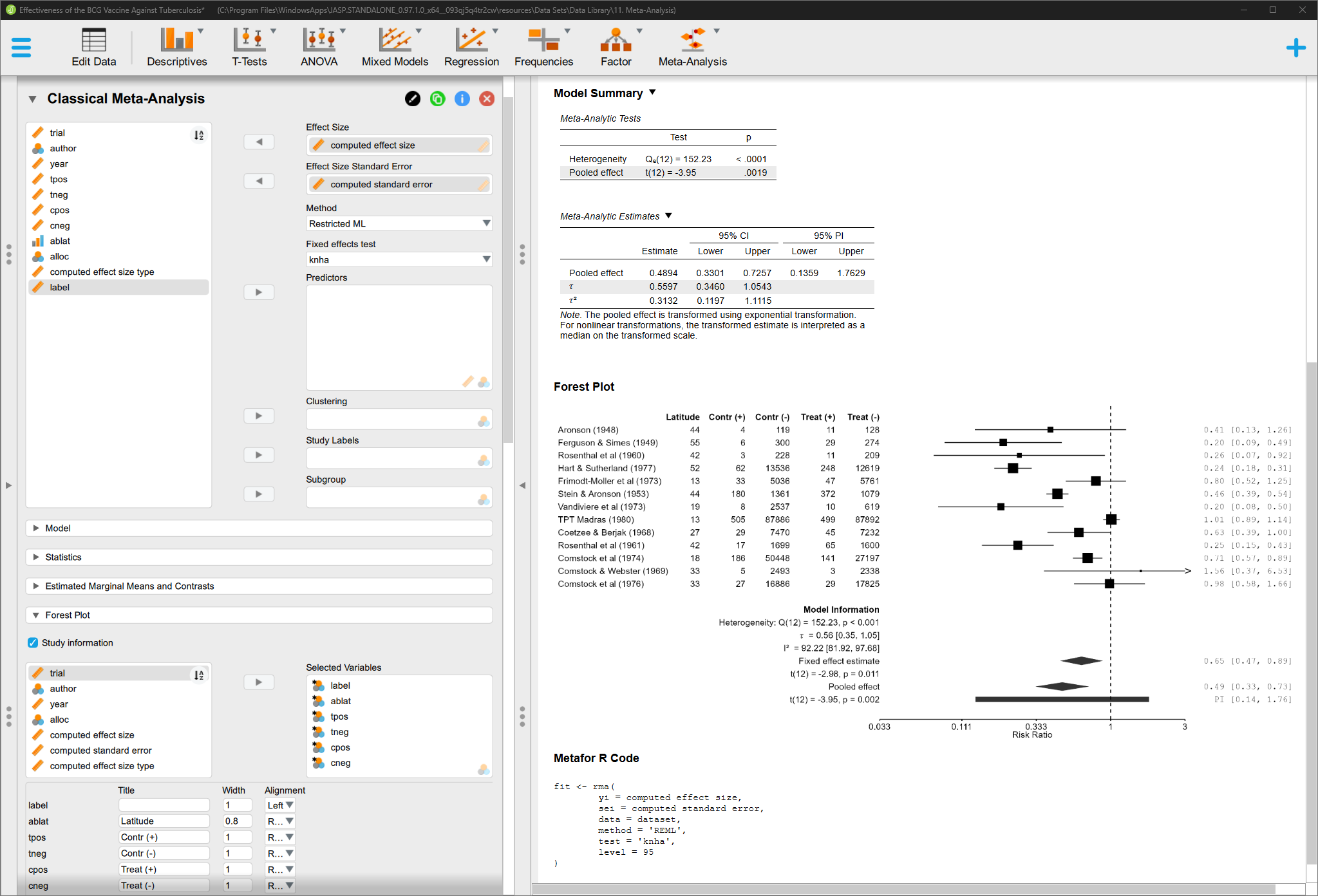}
    \caption{Random-Effects Meta-Analysis With Forest Plot}
    \label{fig:meta-analysis}
\end{figure}

With the variable input set as in the left panel of Figure~\ref{fig:meta-analysis}, the `Meta-Analytic Tests' and `Meta-Analytic Estimates' tables appear in the `Model Summary' output section (right output panel of Figure~\ref{fig:meta-analysis}). In this example, the tests reject the null hypothesis that the pooled effect is zero and the null hypothesis of no between-study heterogeneity. Estimates are by default reported on the input scale (i.e., the log risk ratio). To present pooled estimates and confidence/prediction intervals on the risk ratio scale instead, we set the `Exponential' option in the `Transform effect size' dropdown located in the `Statistics' section. This transformation propagates to all effect size estimates in the forest plot, bubble plots, and estimated marginal means.

Options within the `Statistics' section allow users to obtain additional information about the fitted model, such as relative heterogeneity measures I\textsuperscript{2} and H\textsuperscript{2}, fit measures (e.g., log likelihood, AIC, and BIC), and confidence and prediction intervals of different levels. In addition, the `Diagnostics' section presents comprehensive model diagnostics, including casewise diagnostics \citep{viechtbauer2010outlier, viechtbauer2021model}, a profile likelihood plot \citep{raue2009structural}, the Baujat plot \citep{baujat2002graphical}, and the residual funnel plot. Finally, the `Advanced' section contains additional settings that allow advanced users to fully control the \texttt{metafor::rma.uni} function by changing the optimization settings, fixing between-study heterogeneity, extending the \texttt{R} function call directly, and applying permutation tests \citep{gollmann1999valid, good2009permutation}. Importantly, the `Advanced' section also allows the \texttt{R} code that corresponds to the fitted model to be generated via the `Generate metafor R code' option. The resulting output, shown at the bottom of Figure~\ref{fig:meta-analysis}, can be especially helpful when teaching users of various levels or when attempting a graceful transition to \texttt{R}.

\subsection{Forest Plot}

Forest plots are the most commonly used visualization in meta-analyses and can be created in the `Forest Plot' section. The forest plot can include three components that are toggled independently: `Study information', `Estimated Marginal Means', and `Model Information'. To produce the forest plot shown in Figure~\ref{fig:meta-analysis}, we select the `Study information' component and specify `Selected Variables' to define study information to be shown on the left side of the study information panel (including information about column titles, centering, etc.). We also select the `Model Information' panel and specify the desired information to be included in the panel (e.g., heterogeneity estimates, fixed effect estimate, etc.). Final styling, such as modifying the relative width of the left, middle, and right sides of the figure (automatic width adjustment is not always perfect), setting the $x$-axis label, or the addition of vertical line(s), is controlled in the `Settings' subsection.

Other notable options in the `Study Information' panel include the ability to show predicted effects based on the estimated model, and the ability to aggregate effect size estimates according to a grouping variable, which is especially useful when individual studies contribute multiple effect sizes. The `Settings' subsection controls the details of the output, such as the number of decimals, whether hypothesis tests and/or estimates are reported on the right or left side of the figure, mapping variables to the effect size estimates' color and shape, and additional behavior when a subgroup analysis is specified.

\subsection{Subgroup Analysis}

The first example is completed by performing a subgroup analysis based on the type of treatment allocation. The subgroup analysis is specified via the `Subgroup' variable input. Assigning the allocation variable into the variable input stratifies all previous outputs (tests, estimates, forest plot) based on the treatment allocation. Furthermore, a new row called `Subgroup differences' is added to the `Meta-Analytic Tests' table with results of a null hypothesis of no differences in the pooled effect across the subgroups. Here, we cannot reject the null hypothesis of no difference between the subgroups ($p = 0.361$, not shown).

In the case of binary or person-time outcomes, the Mantel-Haenszel \citep{mantel1959statistical} and Peto \citep{yusuf1985beta} methods with equivalent input and output options are available under the `Mantel-Haenszel / Peto' menu of the Meta-Analysis module.

\section{Example 2: Writing-to-Learn Interventions}

The second example involves a dataset on the effectiveness of writing-to-learn interventions \citep{bangertdrowns2004effect}, distributed with the \texttt{metadat} \texttt{R} package \citep{metadat}. The dataset contains standardized mean differences and sampling variances\footnote{Users can transform the sampling variance into the standard error in the JASP data editor using the square-root transformation.} comparing academic achievement (e.g., final grades, exam scores) of students who received instructions with increased emphasis on writing tasks (i.e., the experimental group) vs. students who received conventional instructions (i.e., the control group) from 48 studies, together with several covariates. Since the effect size estimates and their standard errors are readily available from the dataset we can skip the effect size calculation step and directly proceed to the analysis. For illustration, we focus on two moderators: the continuous intervention length (in weeks) and a categorical indicator of whether participants received feedback.

\subsection{Meta-Regression}

\begin{figure}
    \centering
    \includegraphics[width=1\linewidth]{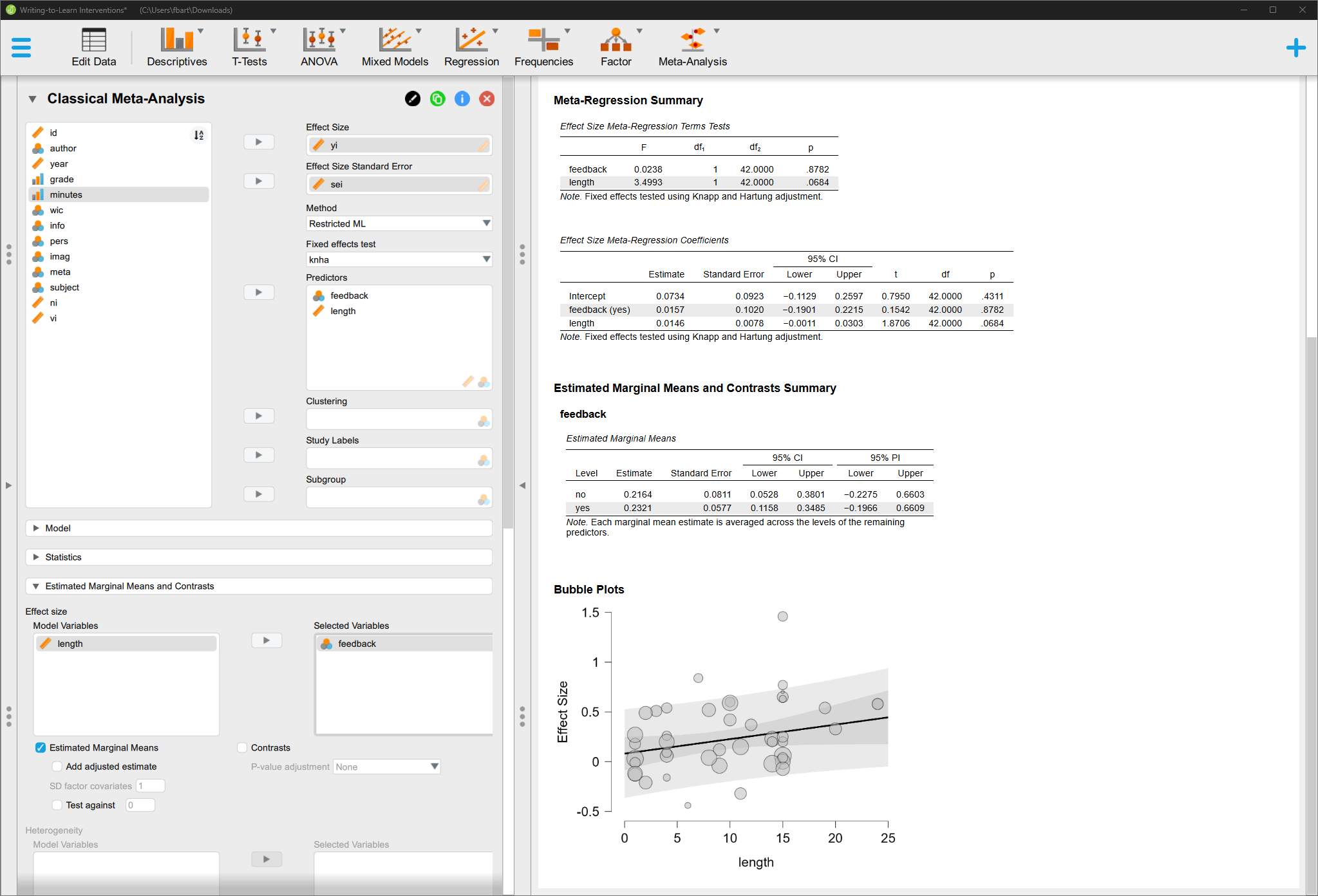}
    \caption{Meta-Regression With Estimated Marginal Means and Bubble Plot}
    \label{fig:meta-regression}
\end{figure}

Meta-regression examines how study-level predictors (moderators, such as the length of the intervention) relate to the effect size while adjusting for other predictors (e.g., whether participants received feedback). In contrast to subgroup analysis, common between-study heterogeneity is estimated from the full dataset after covariate adjustment (unless, as showcased later, a location--scale model is applied).

After setting the required `Effect Size' and `Effect Size Standard Error' variables in the `Meta-Analysis' menu, the meta-regression specification is completed by assigning the intervention `length' and `feedback' variables to the `Predictors' variable input. Continuous and categorical moderators are handled differently by the analysis, and therefore it is important to verify the variable icons: three intersecting circles indicate a categorical variable, whereas the ruler icon indicates a continuous variable. The complete input for our example is shown on the left side of Figure~\ref{fig:meta-regression}.

The corresponding output shows a `Meta-Regression Summary' output section that contains the `Meta-Regression Terms Tests' table. For categorical predictors, the term-specific omnibus test evaluates whether all $p-1$ coefficients equal zero (i.e., a null hypothesis of no differences among the moderator levels). For continuous predictors, the test corresponds to a test of the slope coefficient (i.e., a null hypothesis of zero slope). In our example, neither null hypothesis is rejected at the $\alpha=.05$ level. The `Meta-Regression Coefficients' table then lists the individual meta-regression coefficients.

The omnibus moderation test of all moderators simultaneously is added to the main `Model Summary' output section (not visible in Figure~\ref{fig:meta-regression}). Moderator interaction terms can be added via the `Model' section by selecting multiple `Available Components' while holding the shift key and assigning them to the `Model Terms' variable input.

Missing values in moderators, effect sizes, or standard errors result in list-wise deletion (complete-case analysis). The number of omitted observations (three in our example) is reported below the `Meta-Analytic Estimates' table in the `Model Summary' output section. The JASP Team is currently working on implementing multiple imputation via the \texttt{mice} \texttt{R} package \citep{mice} across all analyses in JASP.

\subsection{Estimated Marginal Means and Contrasts}

The interpretation of meta-regression can be aided by the `Estimated Marginal Means and Contrasts' section. Estimated marginal means (EMMs) are the model-predicted effects at each level of the moderator of interest, with every other moderator held fixed at a representative value: a continuous moderator is set to its mean, and a categorical moderator is averaged over its levels (each level weighted equally). For example, the EMMs for the categorical moderator `feedback' are the predicted effects at its two levels (`no' and `yes'), each evaluated at the mean of the continuous moderator `length'. If a second categorical moderator were present, e.g., `subject' with levels `Math', `Social Studies', and `Biology', then the `feedback' EMMs would still be evaluated at the mean of `length', but would additionally be averaged across the three `subject' levels, so that they reflect no single subject in particular.

Returning to the current example with `feedback' and `age' moderators, Figure~\ref{fig:meta-regression} shows the `Estimated Marginal Means' output table for the `feedback' moderator at levels `no' and `yes'. Furthermore, the EMMs allow users to test whether the predicted effect for each level differs from zero (or any other value of interest). The level-specific estimates can be tested against a null hypothesis of no effect via the `Test against' option with the following value input set to `0' in the left input panel. We find that the estimated effect size at both levels of the moderator is significantly different from zero.

Pairwise differences between the EMMs can be obtained by selecting the `Contrasts' option. Compared to raw meta-regression coefficients, EMMs and their contrasts incorporate moderator interactions\footnote{Note that the \texttt{metafor} function \texttt{pairmat} does not compute EMMs, as it does not average across the levels of the other terms.} and avoid re-leveling of categorical variables. EMMs can also be shown in the `Forest plot' via its corresponding subsection. When a continuous moderator is added to the `Selected Variables' input field, predictions are produced at $\pm$1 standard deviation from the moderator's mean.

The `Estimated Marginal Means and Contrasts' section also provides an `Adjusted estimate' option which calculates the effect for the \emph{unweighted} average across the moderator levels. This differs from the pooled effect in the `Model Summary' output, which is based on the \emph{weighted} average across the moderator levels.

\subsection{Bubble Plot}

The `Bubble Plot' section (not shown in Figure~\ref{fig:meta-regression}) allows users to visualize moderator values against observed effect size estimates as bubbles, with larger bubbles indicating higher precision, together with the estimated meta-regression trend. The bottom-right panel of Figure~\ref{fig:meta-regression} shows a bubble plot of intervention length with a positive, although not statistically significant, increase in the effect size estimate in studies with longer interventions. The trend reflects predictions over the `Selected Variable' input, averaging across the remaining moderators as in EMMs. Specifying `Separate Lines' and `Separate Plots' variables allows users to visualize predicted trends at different levels of other moderators, which is particularly helpful in the presence of interaction effects. The darker band around the regression line depicts the confidence interval for the mean effect; the lighter band depicts the prediction interval for true study effects.

The JASP bubble plots also enable visualization of categorical moderators; instead of a regression line, vertical `box plots' with confidence and prediction intervals are displayed alongside the jittered effect size estimate bubbles.

\subsection{Location--Scale Models}
We conclude the second example by extending the analysis to a location--scale model \citep{viechtbauer2022location}. Location--scale models relax the assumption of a common between-study heterogeneity by including a meta-regression model for the heterogeneity parameter. In other words, standard meta-analysis and meta-regression models assume that the between-study heterogeneity $\tau$ is constant. Location--scale models allow the between-study heterogeneity to vary as a function of the moderators. The scale part of the model (with the location part corresponding to the standard regression on the effect size) is specified in the `Heterogeneity Model (Scale)' subsection of the `Model' section.

The previously estimated `location' meta-regression can be extended to a location-scale meta-regression by assigning the study length moderator to the corresponding (`Heterogeneity Model (Scale)') `Model Terms' variable input. The previous output is extended with appropriate `Meta-regression Terms Tests' and `Meta-regression Coefficients Table' for the scale (Heterogeneity) term. Although the moderation effect on heterogeneity is not statistically significant at the $\alpha = .05$ level, the corresponding bubble plot would show slightly wider prediction intervals for longer interventions.

As with the pooled effect (computed at the weighted average of observed moderator values), the heterogeneity estimates from the location-scale model are also computed at the weighted average of observed moderator values. The adjusted heterogeneity at the unweighted average of moderator levels can be obtained with the `Adjusted estimate' option in the `Heterogeneity' subsection of the `Estimated Marginal Means and Contrasts' section. This section also provides estimated marginal means and contrasts for the heterogeneity regression itself, parallel to those for the effect size regression.

\section{Example 3: Recidivism and Mental Health}

The third example concerns studies on the association between recidivism and mental health \citep{assink2015risk}. This dataset again comes bundled with the \texttt{metadat} \texttt{R} package \citep{metadat}. The dataset contains standardized mean differences and sampling variances comparing recidivism of delinquent juveniles with and without mental health disorders across 17 studies. Each study contributes several effect size estimates for different types of delinquent behavior measured for the same participants, which induces within-study dependence. Accordingly, this dataset allows a demonstration of multilevel models, multivariate models, and cluster-robust standard errors, available from the `Meta-Analysis (Multilevel/Multivariate)' menu of the Meta-Analysis module. These three adjustments are all driven by the same study grouping, but they enter the analysis through three different inputs and address the dependence in three different ways: the multilevel model (the `Level 1' and `Level 2' inputs of the `Random Effects / Model Components' section) captures the similarity among the true effects estimated within a study, the multivariate model (the `Cluster' input of the `Effect Size Variance-Covariance Matrix' section) captures the correlation among the sampling errors of estimates that share participants, and cluster-robust standard errors (the `Clustering' input of the data input section) guard the standard errors against misspecification of that assumed dependence structure. Assigning the study variable through three separate inputs gives the same grouping a different role. Previously described functionality, such as meta-regression, estimated marginal means and contrasts, bubble plots, etc. can be applied to this example as well (apart from scale moderation).

\subsection{Multilevel Models}

\begin{figure}
    \centering
    \includegraphics[width=1\linewidth]{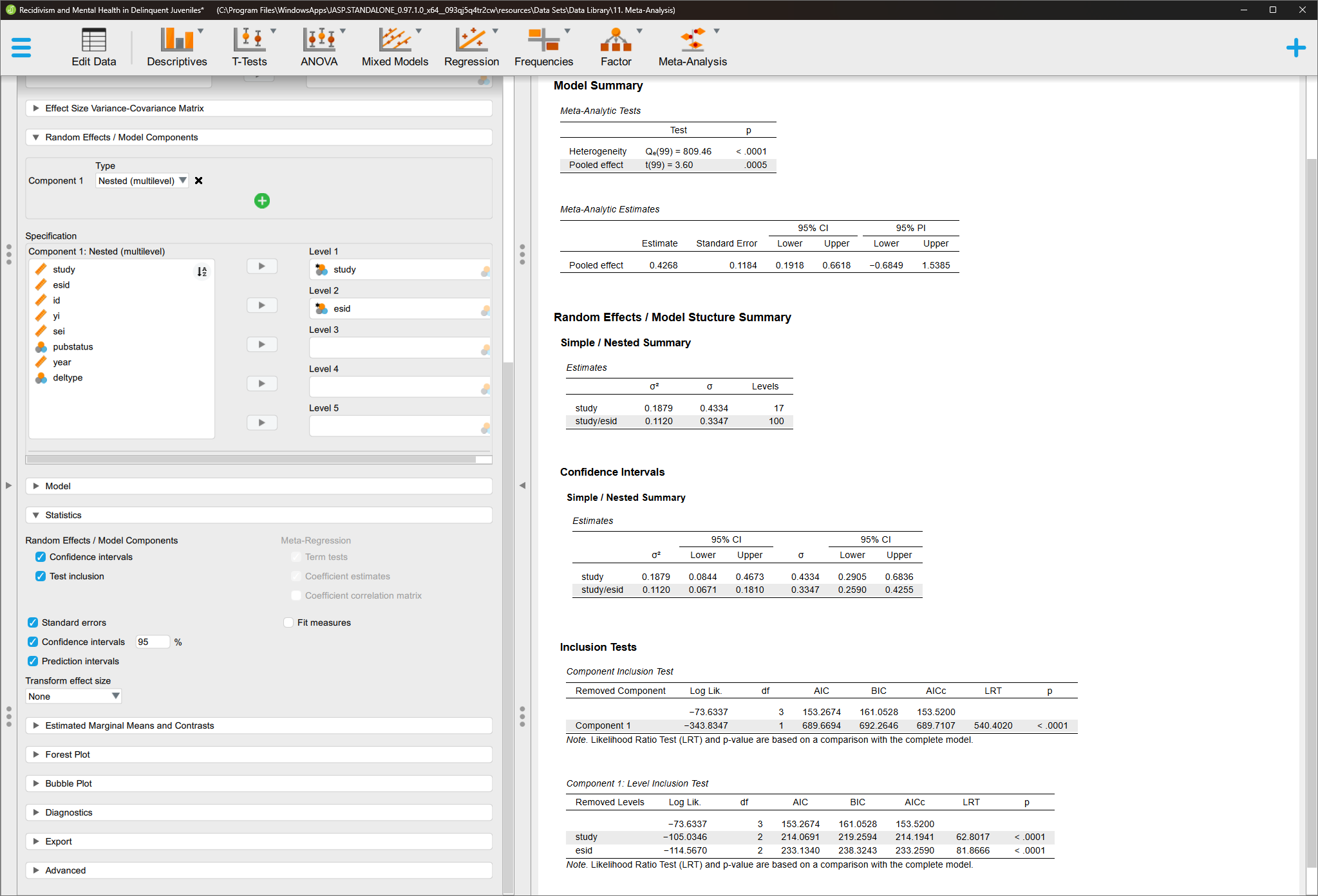}
    \caption{Multilevel Meta-Analysis and Heterogeneity Components Inclusion Tests}
    \label{fig:multilevel}
\end{figure}

If individual effect size estimates are not independent, a simple random-effects model can inflate the precision of the effect size estimate, resulting in overly optimistic $p$-values and poor confidence interval coverage. One source of dependence is that estimates within the same study tend to be more similar than estimates across studies (e.g., due to shared populations, manipulations, or measures). For instance, in the current example each study provides multiple estimates of recidivism; a study with samples from a population with an extensive criminal history will lead to systematically different recidivism estimates than a study with samples from a population with a short criminal history. Multilevel models can account for this dependence by introducing separate heterogeneity components for each ``level'' (i.e., study-level and estimate-level) to account for the estimates-within-studies nesting dependency \citep{konstantopoulos2011fixed, nakagawa2012methodological}.

After assigning the usual `Effect Size' and `Effect Size Standard Error' variables, the multilevel model is specified by setting the `Nested (multilevel)' component in the `Random Effects / Model Components' section. In the `Specification' subsection depicted in the left panel of Figure~\ref{fig:multilevel}, we define `Level 1' as the higher nesting level (study) and `Level 2' as the lower nesting level (estimates). After completing the model specification this way, the output is then extended with a `Random Effects / Model Structure Summary' section shown in the upper right output panel of Figure~\ref{fig:multilevel}, which reports the study-level ($\tau_\text{study}$) and estimate-level ($\tau_\text{study/esid}$) heterogeneity estimates; note that these estimates are no longer summarized in the `Model Summary' section. Confidence intervals for the heterogeneity components, often requiring additional extensive calculation, can be requested via the `Confidence intervals' option in the `Random Effects / Model Components' subsection of the `Statistics' section.

The `Test inclusion' option in the `Statistics' subsection assesses whether the specified random-effects structure improves model fit, and it does so at two levels of granularity. A model may contain one or more random-effects components (the terms listed in the `Random Effects / Model Components' section), and a nested component may in turn span several levels (here, `study' and `estimate'). At a coarse level of granularity, the `Component Inclusion Test' table compares the full model against a model with one entire component removed  (or their combinations); when several components are specified, each is dropped in turn so that its contribution can be judged on its own, which is where this test is most informative. In the present example only a single nested component is specified, so removing it leaves a fixed-effect model. The test rejects the null hypothesis of equivalent fit, suggesting that a random-effects structure is preferred over a model without one.

At a finer level of granularity, and only when a nested component spans multiple levels, the `Component 1: Level Inclusion Test' table compares the full model against models in which individual levels (or their combinations) are dropped. In our example, dropping either the study-level or the estimate-level significantly worsens the fit, so both nesting levels are warranted: the two-level structure is preferred over the simpler single-level alternatives that retain only estimate-level or only study-level random effects.

The `Random Effects / Model Structure Summary' section allows the specification of a wide range of random effects available through the \texttt{metafor::rma.mv} function, including random slopes, structured, autoregressive, and spatial random effects, as well as phylogenetic models, some of which are showcased in the JASP Data Library or under the `Specifying Random Effects' heading at \url{https://wviechtb.github.io/metafor/reference/rma.mv}.

\subsection{Multivariate Models}

\begin{figure}
    \centering
    \includegraphics[width=1\linewidth]{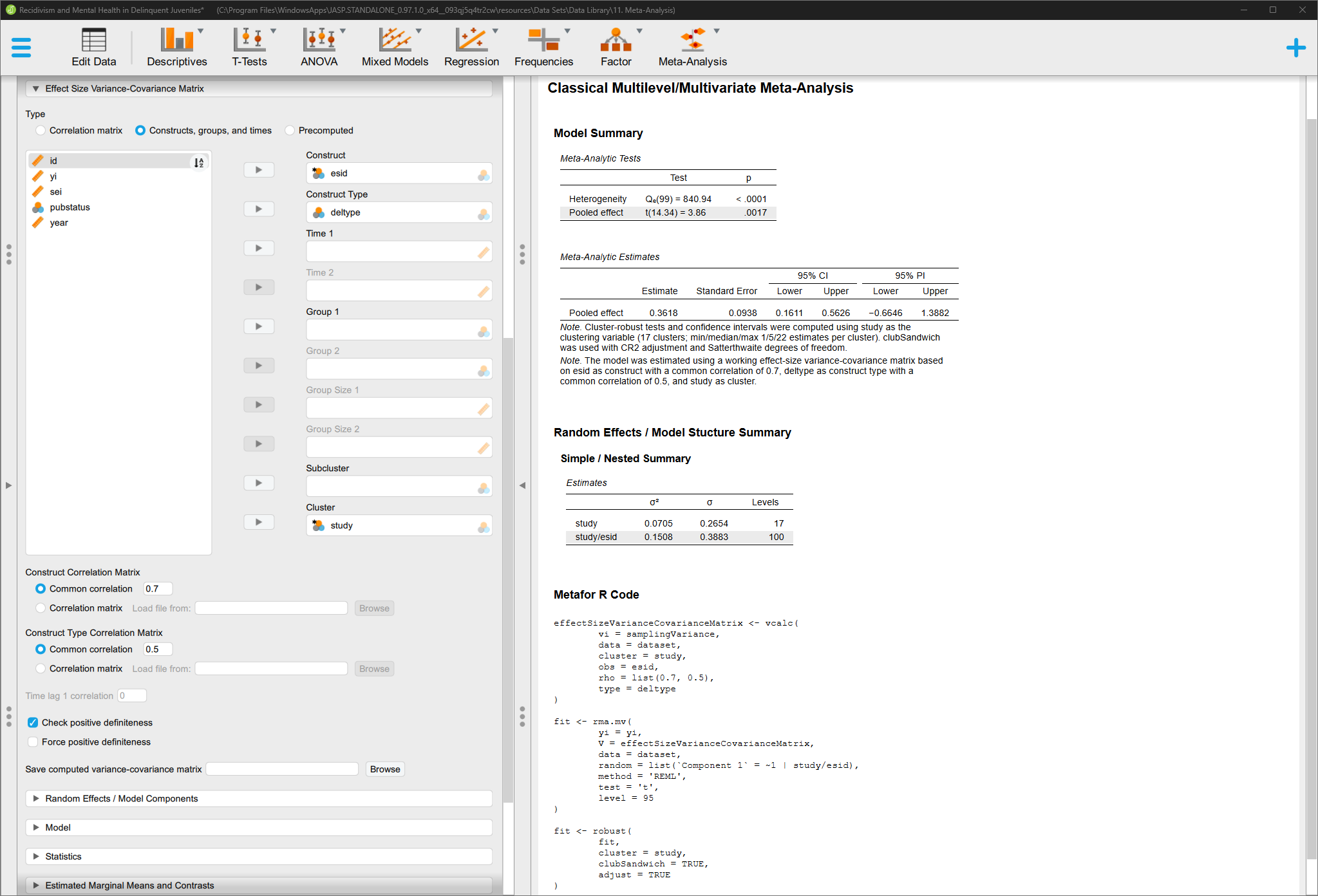}
    \caption{Multilevel and Multivariate Meta-Analysis With Cluster-Robust Standard Errors}
    \label{fig:multivariate}
\end{figure}

Another source of dependence is that repeated assessments on the same participants (across time points or measures) produce correlated estimates that do not contribute fully independent information. For instance, the current example features measures of recidivism in different types of delinquent behavior (e.g., aggression, bullying, stealing)\footnote{The exact recidivism measures are not reported in the original article: we input generic delinquent behavior types for the clarity of the illustration.} from the same groups of participants. Ignoring this multivariate dependency can again result in overly optimistic $p$-values and poor confidence interval coverage \citep{kalaian1996multivariate, houwelingen2002advanced}.

Multivariate models are generally used to account for this particular dependence. For their application these models require the specification of the between-estimate dependencies via the effect-size variance--covariance matrix. This can be done in the `Effect Size Variance-Covariance Matrix' section (see the left input panel of Figure~\ref{fig:multivariate}), which uses the \texttt{metafor::vcalc} function in the background.

In our example, the variance--covariance matrix is built from three variable inputs. The `Cluster' input identifies sets of estimates that may be correlated; estimates in different clusters are assumed independent, and here each study forms its own cluster because studies report on different participants. Within a cluster, the `Construct' input identifies the individual measurements (recidivism in different delinquent behaviors, such as aggression, bullying, and stealing), and the `Construct Type' input groups those measurements into broader categories (here, overt and covert behavior). Specifying both inputs is useful when measurements of the same type are expected to correlate more strongly than measurements of different types: aggression (overt) and bullying (overt) recidivism should correlate more strongly than aggression (overt) and stealing (covert) recidivism. The strength of these correlations is supplied by the user rather than estimated from the data through the `Construct Correlation Matrix' option for measurements sharing a `Construct Type' ($r = 0.70$ in our example) and the `Construct Type Correlation Matrix' option for measurements of different `Construct Type' ($r = 0.50$).

Once these inputs are set, the output updates to reflect the multivariate adjustment (the right panel of Figure~\ref{fig:multivariate}, which also includes the cluster-robust standard errors described next). In our example, the adjustment mostly reduces the study-level and estimate-level heterogeneity estimates: part of the within-study dependence is now attributed to the assumed correlation between sampling errors, so less of it needs to be absorbed by the random-effects variances.

The specification of the variance--covariance matrix can be challenging and error-prone, and therefore we recommend users consult the \texttt{metafor::vcalc} documentation at \url{https://wviechtb.github.io/metafor/reference/vcalc.html} and then use `Show Metafor R Code' in the `Advanced' section to verify the matrix specification. The interface also supports per-level correlation matrices for all `Construct' or `Construct Type' pairs (see the JASP Help File for formatting), and allows dependencies from repeated measures across time or from a shared comparison group via the corresponding variable inputs. Alternatively, correlations between two variables can be supplied directly via the `Correlation matrix' option, or the entire variance-covariance structure can be provided via the `Precomputed' option. Finally, a computed variance-covariance matrix can be saved using the `Save computed variance-covariance matrix' option, manually edited, and reloaded via the `Precomputed' option.

\subsection{Cluster-Robust Standard Errors}

The example concludes with an adjustment that yields cluster-robust standard errors \citep{tipton2015smallsample}. Cluster-robust standard errors are obtained by assigning the study indicator to the `Clustering' input in the data input section (the first, pre-opened section), a separate input from the `Cluster' variable used to build the variance--covariance matrix. After the `Clustering' variable is set, the output in the right panel of Figure~\ref{fig:multivariate} incorporates the adjustment. With the multilevel, multivariate, and cluster-robust adjustments all in place, a statistically significant difference in recidivism between juveniles with and without mental health disorders remains: the pooled effect size estimate is $d$ = 0.36, 95\% CI [0.16, 0.56], $t$(3.9) = 3.86, $p$ = 0.002.

By default, cluster-robust standard errors use the small-sample adjustment from the \texttt{clubSandwich} \citep{clubSandwich} \texttt{R} package. The cluster-robust standard errors settings can be further modified in the `Advanced' section.

\section{Concluding Comments}

This manuscript summarizes the functionality of the classical Meta-Analysis module in JASP. The software is free, open-source, and makes state-of-the-art meta-analytic methodology available within a couple of mouse clicks. This means that analysts, teachers, and students can focus their efforts on choosing the appropriate methodology and interpreting its results, rather than devoting considerable time to writing and debugging relatively complex analysis scripts. JASP also generates \texttt{R} code snippets for the underlying analytic functionality which can facilitate teaching and learning. Finally, the reproducible and easy-to-annotate analysis files directly render on the Open Science Framework, further facilitating reproducibility and data-sharing practices.

This manuscript covered effect size computation, funnel plots, random-effects meta-analysis, forest plots, subgroup analysis, meta-regression, estimated marginal means and contrasts, bubble plots, location--scale models, multilevel models, multivariate models, and cluster-robust standard errors, built using the \texttt{metafor} \citep{metafor} \texttt{R} package. The Meta-Analysis module also contains additional tools not discussed here, including Bayesian meta-analytic methodology mirroring the classical features (presented in Part II; \cite{bartos2025bayesian}), publication bias adjustments \citep{bartos2020adjusting}, meta-analytic generalized mixed-effects models \citep{turner2000multilevel}, meta-analyses of prediction model performance \citep{debray2019framework}, meta-analytic SEM and SEM-based meta-analysis \citep{metaSEM}, penalized meta-analysis \citep{van2023selecting}, effect size aggregation, and risk-of-bias visualizations \citep{robvis}. A full list of the currently implemented features is available at \url{https://github.com/jasp-stats/jaspMetaAnalysis/}. Features such as network meta-analysis \citep{dias2018network} are scheduled for later implementation.

By providing classical meta-analytic capabilities through a user-friendly interface, the JASP Meta-Analysis module makes meta-analytic methods available to a wider audience, and hopefully facilitates the adoption of state-of-the-art methodology across a broad range of scientific disciplines.


\paragraph{Acknowledgments}
František Bartoš is grateful to the JASP Team for reviewing the Meta-Analysis module pull requests and developing new features necessary to implement the module. 

\paragraph{Funding Statement}
This work is supported by ERC grant 101201482, BAYESIAN BOOST, awarded to EJ Wagenmakers. Views and opinions expressed are those of the authors only.

\paragraph{Competing Interests}
František Bartoš and Eric-Jan Wagenmakers declare their involvement in the open-source software package JASP (\url{jasp-stats.org}), a non-commercial, publicly funded effort to make Bayesian statistics accessible to a broader group of researchers and students. In addition, František Bartoš and Eric-Jan Wagenmakers are involved in JASP Services BV, a company that supports organizations who wish to adopt JASP, with a focus on applications in quality control.

\paragraph{Data Availability Statement}
JASP can be downloaded from \url{https://jasp-stats.org/download/} with the source code available at \url{https://github.com/jasp-stats/}. Annotated example analysis files are available at \url{https://osf.io/mb56t/}.  

\paragraph{Ethical Standards}
The research meets all ethical guidelines, including adherence to the legal requirements of the study country.

\paragraph{Author Contributions}
Conceptualization: F.B. Methodology: F.B.; W.V. Data curation: F.B.; W.V. Data visualisation: F.B. Writing original draft: F.B. All authors approved the final submitted draft.


\bibliographystyle{biometrika}
\bibliography{bib/all.bib, bib/software, bib/unpublished} 

\end{document}